\documentclass[prl,twocolumn,english,longbibliography,superscriptaddress]{revtex4-1}

\usepackage{amsmath}
\usepackage{amssymb}
\usepackage{amsthm}

\usepackage{amsbsy}
\usepackage{amstext}
\usepackage{graphicx}
\usepackage{xcolor}
\usepackage{float}
\usepackage{soul}
\usepackage{wasysym}

\usepackage{mathtools}

\makeatletter
\usepackage{amsfonts}
\usepackage{dcolumn}
\usepackage{bbold,bm}

\def\bra#1{\mathinner{\langle{#1}|}}
\def\ket#1{\mathinner{|{#1}\rangle}}
\def\inner#1#2{\mathinner{\langle{#1}|{#2}\rangle}}
\def\braket#1{\mathinner{\langle{#1}\rangle}}

\def\re{\mathrm{Re}\,}
\def\im{\mathrm{Im}\,}

\def\nodag{^{\vphantom{\dagger}}}

\def\note#1{{#1}} 

\makeatother

\begin{document}

\title{Quantum Noise Detects Floquet Topological Phases}

\author{M. Rodriguez-Vega}
\affiliation{Department of Physics, Indiana University, Bloomington, Indiana 47405, USA}

\author{H. A. Fertig}
\affiliation{Department of Physics, Indiana University, Bloomington, Indiana 47405,
USA}

\author{B. Seradjeh}
\affiliation{Department of Physics, Indiana University, Bloomington, Indiana 47405, USA}
\affiliation{Max Planck Institute for the Physics of Complex Systems, N\"othnitzer Stra\ss e 38, Dresden 01187, Germany}

\begin{abstract}
We study quantum noise in a nonequilibrium, periodically driven, open system attached to static leads. Using a Floquet Green's function formalism we show, both analytically and numerically, that local voltage noise spectra can detect the rich structure of Floquet topological phases unambiguously. Remarkably, both regular and anomalous Floquet topological bound states can be detected, and distinguished, via peak structures of noise spectra at the edge around zero-, half-, and full-drive-frequency. We also show that the topological features of local noise are robust against moderate disorder. Thus, local noise measurements are sensitive detectors of Floquet topological phases.
\end{abstract}


\maketitle


\emph{Introduction}.---%
Topological phases of matter are characterized by bulk topological invariants and, via the bulk-boundary correspondence, also by the appearance of topological boundary states (TBSs). Canonical examples of such phases are provided by Chern insulators characterized by bulk integer Chern numbers and chiral edge states, and time-reversal topological insulators characterized by $\mathbb{Z}_2$ indices and counter-propagating helical surface modes~\cite{HasKan10a,QiZha11a}. In equilibrium, the bulk invariant manifests itself in quantized transport coefficients, such as Hall or spin-Hall conductance. In a multi-terminal geometry, this is equivalently understood in terms of TBSs connecting the leads.

Recent theoretical~\cite{OkaAok09a,Gu11,JiaKitAli11a,KunSer13a,CayDorSim13a,RudLinBer13a,KunFerSer14a,NatRud15a,FarPer15a,KunFerSer16a,TitBerRud16a,RodSer17a,Klin16,Klin17} and experimental~\cite{RecZeuPlo13a,WanSteJar13a,JotMesDes14a} progress has uncovered the possibility of engineering topological phases in a system periodically driven at frequency $\Omega$, where topology is characterized, through Floquet theory, by bulk invariants in the \emph{quasienergy} spectrum and, correspondingly, by the appearance of steady-state Floquet TBSs (FTBSs) in the Floquet zone $[-\Omega/2,\Omega/2]$. These Floquet topological phases have a richer structure than their equilibrium counterparts; for example, in addition to ``regular'' FTBSs at the Floquet zone center with the same period as the drive, they can also host ``anomalous'' FTBSs at the Floquet zone edge with twice the period of the drive. A number of studies have connected the Floquet topological invariants and the corresponding FTBSs to observable quantities. For example, a quantized Floquet sum-rule was obtained~\cite{KunSer13a} for conductance summed over terminal biases spaced by integer multiples of $\Omega$ (we are setting $\hbar=1$). The presence of FTBSs in a disordered driven system has also been connected to a generalized bulk magnetization density~\cite{NatRudLin17a}. While these connections expound observable effects of Floquet topology in principle, they do not necessarily lend themselves to experimental detection. Thus, the problem of detecting a Floquet topological phase remains of interest.

\begin{figure}
\begin{center}
	\includegraphics[width=3.4in]{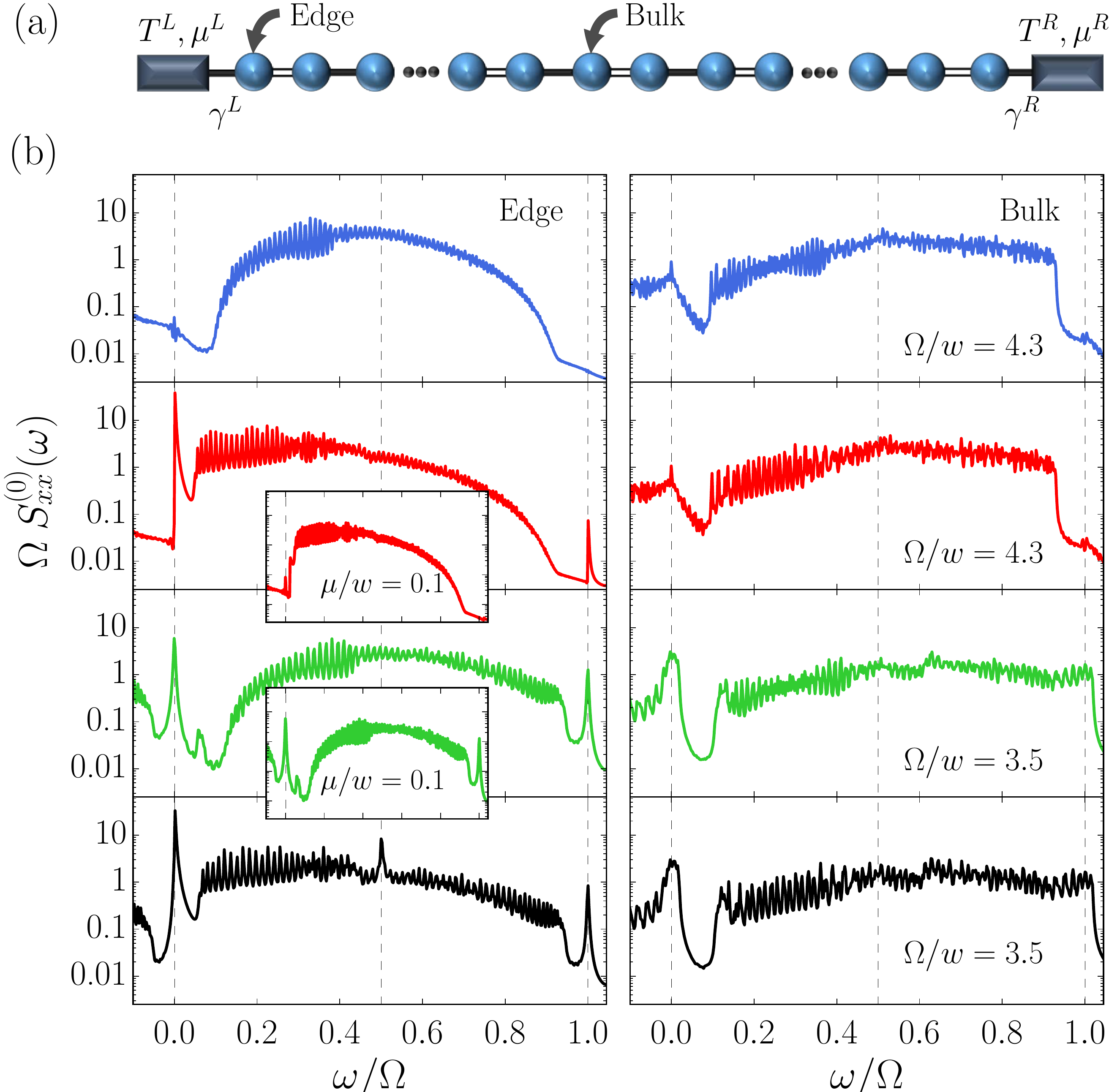}
\end{center}
	\caption{(a) A sketch of the geometry considered for quantum noise. The  system, here a one-dimensional Su-Schrieffer-Heeger chain, is attached to leads at its edges; noise is measured locally at the edge and the bulk. (b) Local voltage noise spectrum, $S_{xx}^{(0)}(\omega)$, of Floquet phases in driven SSH model with the hopping modulation $\delta(t) = \delta_0+\delta_1\sin(\Omega t)$. The left (right) panels show the local noise at the edge, $x=1$ (bulk, $x=50$) of a chain with 100 sites. The top, top middle, bottom middle, and bottom panels are four distinct Floquet phases with, respectively, no, one regular, one anomalous, and both types of Floquet topological bound states at the edge. The dashed vertical lines indicate probe frequencies $\omega=0,\Omega/2$ and $\Omega$. The parameters are: $\delta_0=-0.1$ (top, bottom middle), $\delta_0=0.1$ (top middle, bottom), $\delta_1/w = 0.4$, $\gamma^L=\gamma^R= 10^{-2}w$, $T^L=T^R=10^{-4}w$, and the leads are unbiased in the main panels. The insets show the spectra for a lead bias $0.1w$ and have the same range as the main panels.}
	\label{fig:noise}
\end{figure}

\note{In this Rapid Communication, we show that FTBSs can be detected via noise measurements. We show that quantum noise in a driven system attached to static leads probes the quasienergy excitation spectrum.
Thus, local voltage noise spectrum at the boundary of the system detects both the regular and anomalous FTBSs through peak structures appearing at noise frequencies $\omega=0, \Omega/2$ and $\Omega$. These peaks are absent in the trivial phase and in the bulk, and are robust to static potential disorder. Furthermore, their behavior with respect to lead bias provides unique signatures of their topological origin. A summary of our results is presented in Fig. \ref{fig:noise}.
}

\note{Our proposal differs from most noise studies of electronic systems in its 
focus on voltage rather than current, for which the latter is largely 
irrelevant in the system we consider. Voltage noise resulting from particle number fluctuations can be measured in solid-state as well as cold-atom realizations.
Additionally, by attaching the system to static leads we avoid the problem of heating to infinite temperature, and a featureless noise spectra, in generic driven systems~\cite{DAlRig14a,LazDasMoe14b}. In what follows we first derive a general analytical expression for voltage noise in a Floquet system, and then apply
this formalism to a particular system that hosts FTBSs, the driven Su-Schrieffer-Heeger (SSH)
model ~\cite{SuSchHee79a,RodSer17a}.}

\emph{Model}.---%
As a concrete realization of a Floquet topological system we focus on the driven SSH model, 
which, while simple, exhibits all the relevant
Floquet topological phases. The Hamiltonian is defined on a one-dimensional lattice by
$
\hat H (t) = \sum_x{[w - (-1)^x \delta(t)] \hat c^\dagger_{x+1} \hat c \nodag_x+\text{h.c.}},
$
where $\hat c^\dagger_x$ creates a fermion at lattice site $x$,
$w$  is the unmodulated hopping amplitude and $\delta(t)=\delta(t+2\pi/\Omega)$ is a temporally
periodic hopping modulation.  Using Floquet's theorem,
the Schr\"odinger equation may be written as $[\hat H(t) - i\partial/\partial t]\ket{u_\alpha(t)} = \epsilon_\alpha\ket{u_\alpha(t)}$, where $\ket{u_\alpha(t)} = \ket{u_\alpha(t+2\pi/\Omega)}$ are periodic Floquet steady states with quasienergy $\epsilon_\alpha$.
The static system has two distinct phases: a topological one
for $\delta/w>0$ and a trivial one for $\delta/w<0$, characterized respectively by the presence and absence of
solutions representing TBSs at each edge.
By contrast, the Floquet system has more distinct topological phases; in what follows, 
we focus on four distinct Floquet phases, each with one or no regular and/or anomalous FTBS.

\emph{Noise in Floquet formalism}.---%
Quantum noise in a local observable $\hat O_x(t)$ at time $t$ (in the Heisenberg picture) is defined as
\begin{equation}\label{eq:qnoise}
S_{xy}(t,s) = \frac12 \braket{\{ \hat O_x(t), \hat O_{y}(s) \}} - \braket{\hat O_x(t)}\braket{\hat O_{y}(s)},
\end{equation}
where $\{\cdot,\cdot\}$ is the anticommutator and $\braket{\cdots}$ is the expectation value
with respect to steady states defined by the leads, discussed below.
In our driven systems, the dynamics is given by a periodic Hamiltonian $\hat H(t)=\hat H(t+2\pi/\Omega)$, so the noise is a function of $\tau=t-s$ and is periodic in $t$: $S(t,\tau) = S(t+2\pi/\Omega,\tau) = \frac1{2\pi} \int d\omega  \sum_m  e^{i\omega \tau + i m\Omega t} S^{(m)}(\omega)$, where
$S^{(m)}(\omega)$ defines the Floquet noise spectrum.

\note{We focus on voltage noise, characterized by the spatiotemporal correlations in the number operator $\hat c_x^\dagger \hat c\nodag_x$, where $\hat c_x^\dagger$ creates a quasiparticle in  state $\ket{x}$ (here position). To calculate the expectation values in Eq.~(\ref{eq:qnoise}),
one must specify a density matrix~\cite{DAlRig14a,LazDasMoe14b,IadCha15a,DehMit16a,MoeSon17a} which sets the occupation of the Floquet bands, and therefore plays a defining role in the accessible topological properties of the system. We assume that the system is attached to external static leads, as shown in Fig.~\ref{fig:noise}(a), with a thermal distribution of electrons impinging on the system \cite{Gu11}. This contact with external reservoirs guarantees the
system will not heat to infinite temperature.
We employ a Floquet Green's function approach 
~\cite{KunSer13a,KunFerSer14a,FarPer15a,KunFerSer16a,KohLehHan05a,Note1}
to evaluate the expectation values of Eq. (\ref{eq:qnoise}). The details of our calculation are presented in~\cite{supp}, with
the result,
\begin{align}
S^{(m)}_{xy} (\omega)
&= 2\pi \re \sum_{{\substack{kln \\ \lambda \kappa}}}
\int d \nu
\bra{y}W^{\lambda}_{k+n,n+l}(\nu)\ket{x} f^\lambda(\nu)
\nonumber \\
& \times {\bra{x}W^{\kappa}_{m+l,k}( \nu+\omega+n\Omega)\ket{y}}
\bar f^{\kappa}( \nu+\omega+n\Omega),	
\label{eq:S}
\end{align}
where $k,l,n,$ are integers,
the Fermi distribution of lead $\lambda$ with chemical potential $\mu^\lambda$ and temperature $T^\lambda$ is
$f^\lambda(\omega) = 1/\left(1+\exp[(\omega-\mu^\lambda)/T^\lambda]\right)$, $\bar f^\lambda = 1-f^\lambda$, and $W^{\lambda}_{n,m}(\omega) = G^{(n)}(\omega) \Gamma^\lambda(\omega) G^{(m) \dagger}(\omega) = W^{\lambda\dagger}_{m,n}(\omega)$, with  $\Gamma^\lambda$ the self-energy due to lead $\lambda$.  The matrix elements of the Floquet Green's function $G^{(n)}(\omega)$ give the amplitude of propagation for a particle at energy $\omega$ dressed with $n$ ``photons,'' carrying $n\Omega$ quanta of drive energy. In the wide-band limit,
i.e. with constant lead densities of states, we have the spectral representation
\begin{equation}\label{eq:Gnw}
G^{(n)}(\omega) =\sum_{k \alpha} \frac{\ket{u^{(k+n)}_\alpha}\bra{\bar u^{(k)}_\alpha}}{\omega - (z_\alpha + k \Omega)},
\end{equation}
where $\ket{u_\alpha^{(k)}}$ is the $k$th harmonic of the periodic Floquet state
$\ket{u_\alpha(t)}$, and $z_\alpha$ is the complex-valued quasienergy of the open system.
Note that the adjoint Floquet states $\bra{\bar u_\alpha}^\dagger$ have complex conjugate quasienergy $\bar z_\alpha$. 
}



\begin{figure}[t]
	\includegraphics[height=0.95in]{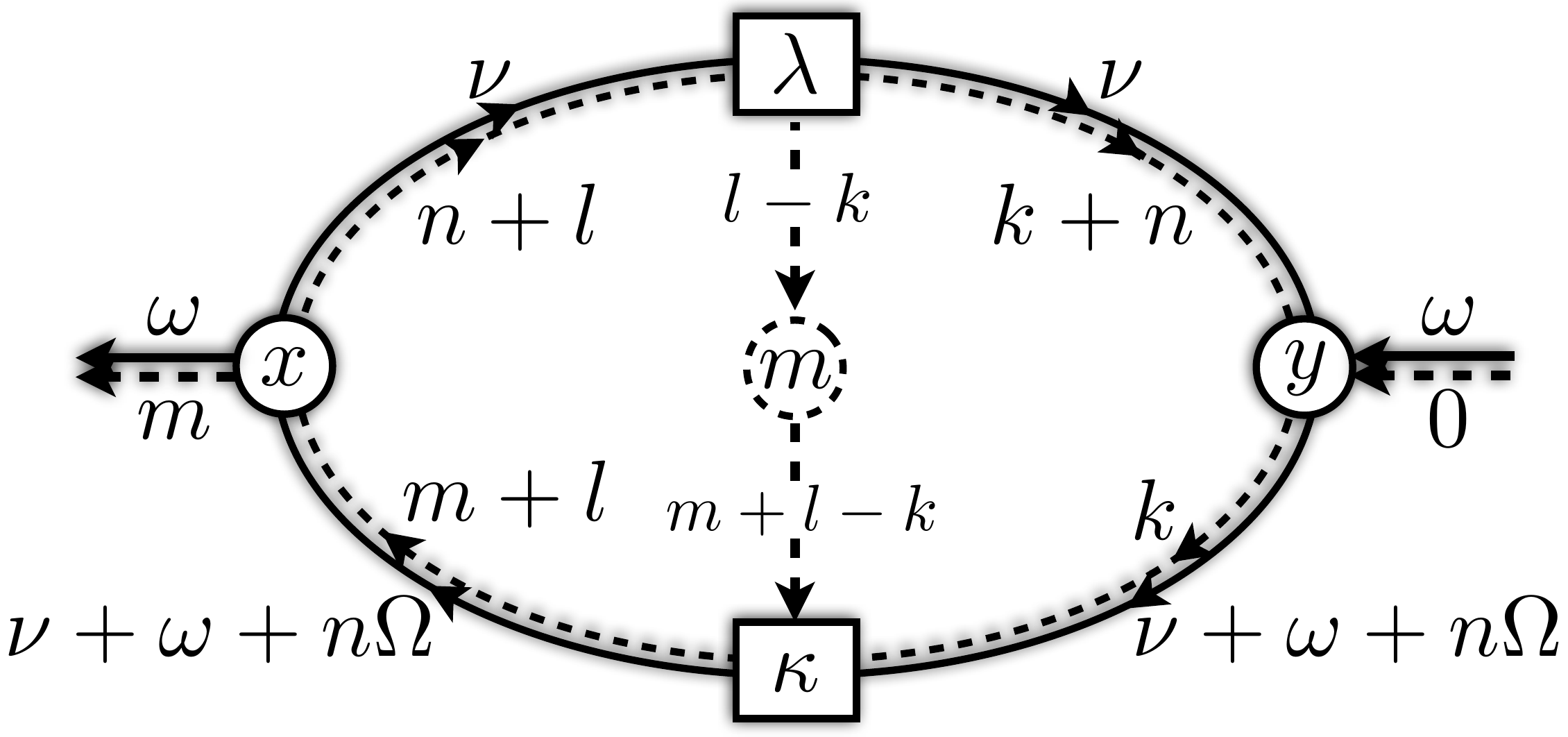}
	\caption{\note{Diagrammatic representation of Eq.(\ref{eq:S}), the Floquet quantum noise $S^{(m)}_{xy}(\omega)$ at probe frequency $\omega$ and $m$ ``photons'' exchanged with the drive across two (position) states $\ket{x}$ and $\ket{y}$. The circular vertices (labeled $x$ and $y$) project to position basis; solid (dashed) lines show particle (photon) propagation in the direction of arrows; square vertices $\lambda$, $\kappa$ represent coupling with  external leads and exchange of a conserved number of ``photons,'' with the net influx of $m$ photons indicated by the dashed circular vertex.}}
	\label{fig:diag}
\end{figure}

\note{Eq. (\ref{eq:S}) can be represented diagrammatically as shown in Fig.~\ref{fig:diag}, with the following rules.} A \note{circular vertex labeled $x$ projects to the state $\ket{x}$}; it is connected to one external line at a given energy and photon number and two dressed particle propagators of a given energy and photon number.
\note{At such a vertex, the outgoing electron energy is increased from the incoming
energy by the incoming external frequency plus the energy exchanged with the drive, equal to $\Omega$ times the {\it net} incoming photon number. Photon number by itself is not conserved at this vertex.}
A \note{square vertex labeled $\lambda$ supplies the tunnel coupling $\Gamma^\lambda$ to lead $\lambda$}; 
it connects one dressed particle to a dressed hole and can exchange a number of photons with the drive. 
Here, the dressed particle energy and photon energy are conserved separately. 
\note{The net influx of $m$ photons is shown by the dashed circular vertex.}
Finally, vertices are connected by lines of propagating dressed particles and holes, represented by $G^\dagger$ and $G$.   This diagram suggests the noise spectrum is understandable in terms of particle-hole
pair fluctuations around the steady state.
\note{We note Eq.~(\ref{eq:S}) is general, and applies for systems in any spatial dimension, type of drive, and coupling with the external leads.}

\note{\emph{Weak-coupling limit}.---%
	While Eq.~(\ref{eq:S}) is valid in general, the physical processes contributing to voltage noise become particularly transparent in the wide-band and weak-coupling limits.
	Then, to lowest order in $\Gamma=\sum_\lambda\Gamma^\lambda$, we can take $\ket{u_\alpha}$ to be the Floquet state of the closed system with quasienergy $\epsilon_\alpha=\re z_\alpha$, $\bra{\bar u_\alpha}=\bra{u_\alpha}$, and $\im z_\alpha \equiv\gamma_\alpha = -\sum_k\bra{u_\alpha^{(k)}}\Gamma\ket{u_\alpha^{(k)}}$.
	Using the spectral form (\ref{eq:Gnw}), we find only the diagonal elements $\bra{u_\alpha^{(k)}}\Gamma^\lambda\ket{u_\alpha^{(k)}}$ contribute significantly to the matrix elements of $W^\lambda$, so that
	\begin{align}
	S^{(m)}_{x y} (\omega) &\approx
	2\pi \re \sum_{{\substack{k l n \\ \lambda \kappa \alpha\beta}}}
	\Upsilon^{{\lambda} (k)}_\alpha \Upsilon^{{\kappa} (l)}_\beta
	f^\lambda( \epsilon_\alpha + k\Omega) \bar f^{\kappa}(\epsilon_\beta + l \Omega )
	\nonumber \\
	& ~~~~\times M_{\alpha\beta,xy}^{(m,n)}
	\delta_{\gamma_{\alpha\beta}}(\omega +\epsilon_\alpha- \epsilon_\beta   + n \Omega ).
	\label{eq:weak}
	\end{align}
	Here,
	$
	\Upsilon^{\lambda(k)}_\alpha = \pi \bra{u_\alpha^{(k)}} \Gamma^\lambda\ket{u_\alpha^{(k)}}/\gamma_\alpha
	$
	is a dimensionless parameter that describes the coupling of the $k$-th Fourier mode of the system with the leads,
	the matrix elements
	$
	M_{\alpha\beta,xy}^{(m,n)} = \sum_{q,p}
	\inner{u^{(n+q)}_\alpha}{x}
	\inner{x}{u^{(m+q)}_\beta}
	\inner{u^{(p)}_\beta}{y}
	\inner{y}{u^{(n+p)}_\alpha}
	$
	account for the projection of the states $|x\rangle, |y\rangle$ into the Floquet basis,
	and
	$\gamma_{\alpha\beta} \approx \max(\gamma_\alpha,\gamma_\beta) + \gamma_0$
	is a small broadening entering the Lorentzian
	$ \delta_{\varepsilon}(z) = 
	(\varepsilon/\pi)/(z^2+\varepsilon^2)$. The delta function makes explicit that
	$S^{(m)}_{xy}(\omega)$ is a measure of the particle-hole \emph{excitation spectrum},
	with a dressed particle and a dressed hole propagating between the positions $x$ and $y$ at energy $\omega$ and with a net loss of $m$ photons to the drive.
	We have included an additional small phenomenological part, $\gamma_0$, to account for other sources of broadening and experimental resolution.}
	
\note{Since in each process the dressed particles and holes can lose or gain photons, one must sum over the amplitudes of all such virtual processes weighted by the appropriate tunnel couplings and matrix elements. 
Thus, it becomes possible to measure the quasienergy excitation spectrum and reveal the presence of FTBSs. This gives rise to noise at frequencies forbidden in a static system, where the only nonzero noise harmonic is $S^{(0)}_{xy}(\omega)$. 
	In an unbiased static system, voltage noise vanishes for frequencies below the particle-hole excitation gap. In particular, the ``shot'' noise at $\omega=0$ vanishes unless there is a resonant (bound) state at the lead chemical potential. This structure can be used to detect static TBSs in equilibrium~\cite{supp}.}
	
\note{In the zero-temperature limit, an additional factor must be included in the summands that correctly accounts for restrictions arising from the step-function Fermi distributions. For small lead bias, this only significantly affects the behavior around $\omega=0$ resulting from $\epsilon_\alpha=\epsilon_\beta=0$. For example, in the static limit, these conditions restrict $\omega>0$ in the zero-temperature limit~\cite{supp}. The main effect of this restriction is to render the peak at $\omega=0$ resulting from a regular FTBS asymmetric, as we discuss below.
}

\emph{Floquet noise spectrum}.---%
We now show that \note{local voltage noise with $x=y$ in the driven system can detect} different types of FTBSs unambiguously. In particular, the structure of the noise spectra near frequencies $\omega=0, \Omega/2$ and $\Omega$ bear unique signatures of FTBSs. For simplicity, we shall assume the system is coupled at its edges to two leads. Very generally, a particle-hole pair between any Floquet steady state and a state in the lead,
dressed with a sufficient number of virtual photons, will contribute to noise at $\omega=0$. While the matrix elements for large virtual photon numbers are quite small, this nevertheless leads to a broad resonance at zero frequency in the bulk.
Some residual zero-frequency noise from these bulk states will persist in the local noise at the edge.

On the other hand, for a topologically nontrivial driven system coupled to unbiased leads, with a regular or
anomalous FTBS, or both, we expect to see a {\it sharp} zero-frequency peak in the local noise measured at the edge, arising from particle-hole excitations among different FTBSs. Similarly, a peak should be seen at $\omega=\Omega$ due to processes involving a single virtual photon.
The peak originating from FTBSs can be distinguished from the broad bulk peak by its behavior with
chemical potential in the lead: in the regular case, it drops sharply in magnitude when the chemical potential
moves away from zero energy.  The analogous peak for the anomalous case
remains unchanged for small biases and only drops sharply when
 the chemical potential passes through $\pm\Omega/2$.
In the bulk, this behavior is completely absent.

When the driven system hosts both regular and anomalous FTBSs, their simultaneous presence
announces itself through the noise spectrum near $\omega=\Omega/2$, since the inter-level particle-hole spectrum at the edge now has an excitation precisely at this energy. Indeed, this peak is robust against a whole set of parameter variations, including disorder and other perturbations, so long as the FTBSs continue to exist. Together with its absence in the local noise spectrum in the bulk, this peak provides an unambiguous detection signal for this intrinsically nonequilibrium topological phase.

\emph{Numerical results}.---%
In Fig.~\ref{fig:noise}(b), we plot the local noise at the edge and the bulk, obtained using the weak-coupling approximation, Eq.~(\ref{eq:weak}), for the driven SSH chain attached to unbiased leads at its edges. We have numerically checked that this approximation accurately reproduces the results of the full expression, Eq.~(\ref{eq:S}), but Eq.~(\ref{eq:weak}) allows for simulation of considerably larger systems. As expected, the trivial phase shows residual zero-frequency noise. However, in all the topological phases prominent zero- and full-frequency peak structures appear at the edge. For the phase with both regular and anomalous FTBSs, an additional peak structure is observed at half-frequency.

We also show, in the insets of the middle two panels, the local noise at the edge for a small bias between the leads. In agreement with our analysis above, the peak structure for the regular (anomalous) FTBS goes away (persists), thus distinguishing the two kinds of Floquet phases. We note that the peaks for the regular (anomalous) FTBS showing an asymmetric (symmetric) shape. For the phase hosting both (the bottom panel), the peak shape shows an intermediate asymmetry. These shapes arise due to restrictions placed on the resonance conditions by the Fermi distributions; a similar asymmetric zero-frequency peak is also observed at the edge of the topological phase of the static system.

\begin{figure}[t]
	\includegraphics[width=3.4in]{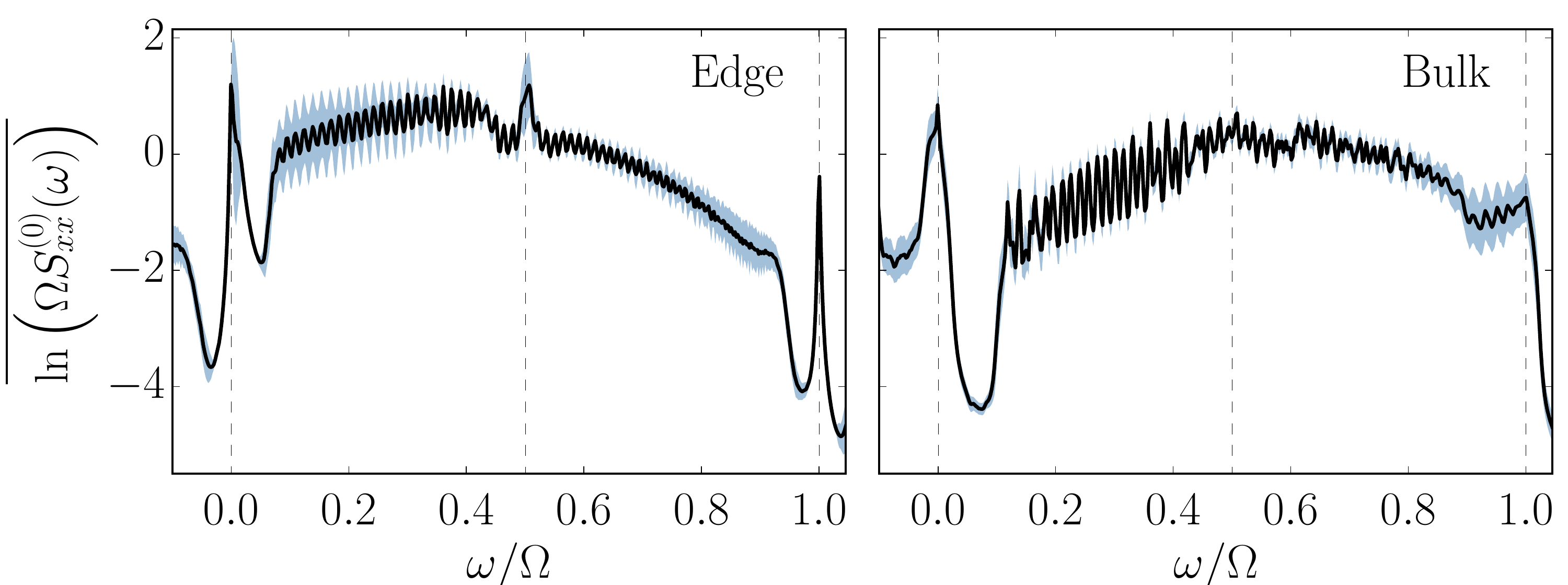}
	\caption{Effect of local potential disorder on local voltage noise, averaged over 200 disorder realizations with strength $W/w=0.1$. The shaded regions show the standard deviation of the signal. The system parameters are as in the bottom panel of Fig.~\ref{fig:noise}(b).}
	\label{fig:dis}
\end{figure}

\emph{Effects of disorder}.---%
Topological bound states hosted by a topological phase are generically robust against local changes of potential that do not spoil the symmetries protecting them: while their wavefunctions change, they stay bound near an edge. Consequently, they are also robust against local disorder that preserves the relevant symmetries on average. Thus, a detection scheme of a topological phase must also display a degree of robustness against disorder. Indeed, we may expect a more dramatic response to disorder, whereby the topological feature becomes more prominent as the nontopological aspects are suppressed by disorder more rapidly~\cite{KunSer13a}.

In order to study the effects of disorder in quantum noise, we calculated the local noise in the SSH model with local potential disorder, taken to be an uncorrelated, uniformly distributed random variable $V_x\in[-W,W]$. An example of our results is shown in Fig.~\ref{fig:dis}. After averaging over disorder, we observed a smoothening of the non-topological features; for example, the local noise at the edge at frequencies away from multiples of $\Omega/2$ shows significant variation, which, after averaging, result in a smoother profile. The peak structure at multiples of $\Omega/2$ shows reduced variation and remains robust. This robustness allows the identification of topological phases in moderately disordered systems.

\emph{Discussion}.---%
As seen in the bottom two panels of Fig.~\ref{fig:noise}(b), the residual zero-frequency signal in the bulk develops to a more prominent peak than in the upper two panels. This structure reflects the growing intraband quasienergy particle-hole excitations with $n=0$ in Eq.~(\ref{eq:weak}), while the larger frequency features result from interband excitations. We also note that the gap around $\omega=\Omega$ clearly seen in the top two panels closes in the bottom two panels. This can also be understood as arising from intraband excitations with $n=-1$ in Eq.~(\ref{eq:weak}). More generally, with decreasing drive frequency, multi-photon processes become more relevant, so that gaps in the bulk noise spectrum fill in and the signal becomes more or less featureless. However, the noise at the edge remains gapped and the peak structures persist at even these lower drive frequencies. This robustness of the edge noise spectrum reflects the robustness of the FTBS which dominate the observed noise signal.

The numerical calculations we report here have been performed in a one-dimensional system. However, the analytical expressions we obtain for local voltage noise, Eqs.~(\ref{eq:S}) and (\ref{eq:weak}), are valid for any dimension. 
\note{For the SSH model as well as other topological one-dimensional systems, 
the focus on voltage rather than current noise is important because the
FTBSs do not carry current.}
In higher dimensions, the system boundaries allow for different geometries of connecting to leads. For example, one may contact the leads on two extended edges or surfaces of a two- or three-dimensional system~\cite{Vis03a}. These geometric variations would lead to different matrix elements, $W^\lambda$ in Eq.~(\ref{eq:S}) or $\Upsilon^\lambda$ in Eq.~(\ref{eq:weak}). It would be interesting to explore the Floquet noise spectra in these and other multi-terminal geometries.

In conclusion, \note{we derived a general expression for the voltage noise in periodically-driven systems.} We have shown that quantum noise in a periodically driven system \note{attached to static leads} probes the quasienergy excitation spectrum. Thus, local quantum noise can detect Floquet topological bound states at the edge and, in particular, the intrinsically nonequilibrium, anomalous Floquet bound states. The structure of Floquet noise spectra at lower drive frequencies and the effects of dissipation and other bath geometries are interesting problems for the future.

\begin{acknowledgments}
We acknowledge useful discussions with Phil Richerme. This work was supported in part by the US-Israel Binational
Science Foundation grants No. 2014345 (M.R.V. and B.S.) and No. 2016130 (H.A.F), the NSF CAREER grant DMR-1350663 (M.R.V. and B.S.), NSF grants DMR-1506263
and DMR-1506460 (H.A.F.), and the College of Arts and Sciences at Indiana University Bloomington. B.S. and H.A.F. also thank the hospitality of Aspen Center for Physics (NSF grant PHY-1607611), where parts of this work were performed.
\end{acknowledgments}


\renewcommand{\thefigure}{S\arabic{figure}}

\section{Supplemental Material}
In this Supplemental Material, we outline the derivation of the Floquet power spectrum, present
the noise in the weak coupling limit and show numerical results for a static system.

\section{Voltage noise derivation}

The Hamiltonian of the system we consider is given by
\begin{equation}
\hat H(t) = \hat H_{S}(t) + \sum_{\lambda} \hat H_{\lambda} + \sum_{\lambda} \hat T_{\lambda},
\end{equation}
where $\hat H_{S}(t)= \hat H_{S}(t+T)$ is the time-periodic system's Hamiltonian, 
which in general can be written as
$
\hat H_{S}(t) = \sum_{xy} w_{xy}(t) \hat c^{\dagger}_{x}\hat c_y,
$
where $w(t)=w(t+T)$ is a hermitian matrix, and 
$\hat c^{\dagger}_{x} ( \hat c_x)$ creates (annihilates) a fermion in the quantum state $x$
of the system. 
$
\hat H_{\lambda} = \sum_{xy} F^\lambda_{xy} \hat a^{\lambda \dagger}_x \hat a^{\lambda}_y,
$
is the Hamiltonian that described lead $\lambda$, 
where $a^{\lambda \dagger}_x$ creates a fermion in state $x$ on lead $\lambda$.
Finally, $
T_{\lambda} = \sum_{x y} K^\lambda_{x y} \hat a^{\lambda \dagger}_x \hat c_y + h.c.
$ is the tunneling Hamiltonian that describes the coupling between lead $\lambda$ and the system.

The time evolution of the 
creation and annihilation operators is obtained solving the coupled
Heisenberg equations of motion 
$
d \hat a^\lambda_{x}/dt  = i \left[ \hat H(t), \hat a^\lambda_x \right]
$
and
$
d \hat c_x/dt = i \left[\hat H(t), \hat c_x\right].
$
We find that the solutions are given by \cite{kohler2005}
\begin{align}
\hat a^\lambda(t) 
	&= i   G^\lambda (t-t_0) \hat a^\lambda(t_0) + \int_{t_0}^t  G^\lambda (t-t') K^\lambda \hat c(t') dt' \nonumber \\
\hat c(t) 
	&= \sum_n \int \frac{d \omega}{2 \pi} e^{-it(\omega  + n\Omega)} G^{(n)}(\omega) \hat h(\omega).
\label{eq:solns}
\end{align}
%
%
The first term in the expression for $\hat a^\lambda(t)$ represents coherent time evolution of the static leads starting at time $t_0$, where $G^\lambda(t)$ is the propagator in lead $\lambda$. The second term arises from coupling with the  time-dependent system $\hat H_S(t)$. The time evolution of $\hat c(t)$ is dictated by
$\hat h(\omega) = \int e^{i \omega t} \hat h(t) dt$,  
$\hat h(t) = i \sum_{\lambda} K^{\lambda \dagger}  G^\lambda(t-t_0) \hat a^\lambda(t_0) $, 
and the Floquet-Green's function  $ G^{(n)}(\omega) $ is defined as the Fourier transform
\begin{equation}
G(t, t') = \sum_n \int G^{(n)}(\omega) e^{-i \omega(t-t')-in\Omega t}
\frac{d\omega}{2\pi}
\end{equation}
of the two-time Green's function $G(t,t')$, given by the solution of
$
\left[i  \partial/\partial t - w(t) \right] G(t,t') - i \int_0^\infty \Gamma(\tau) G(t-\tau,t') d\tau = \delta(t-t'),
$
where $ \Gamma(\tau) = \sum_{\lambda}\Gamma^\lambda(\tau)$, and
$\Gamma^\lambda(\tau) = -i K^{\lambda \dagger} G^\lambda(\tau) K^\lambda$ arise from  coupling with the leads. \\

In this work, we are interested in the voltage noise, which is proportional to occupation number fluctuations defined by the correlation function
\begin{equation}
	S_{xy}(t,s) = \frac{1}{2}\langle \{\hat n_x(t), \hat n_y(s) \}\rangle - \langle \hat n_x(t) \rangle\langle \hat n_y(s) \rangle,
	\label{eq:noise_definition}
\end{equation}
where $\hat n_x (t) = \hat c_x^{\dagger}(t) \hat c_x(t)$ is
the number operator, $\langle \cdots \rangle $  represents a thermal average of the 
 unperturbed system in the far past at time $t_0 \rightarrow -\infty$
defined by the state of the leads
$
\langle \hat a^{\lambda \dagger}_{\nu}(t_0) 
\hat a^{\lambda' }_{\nu'}(t_0)\rangle  =\delta_{\nu \nu'} \delta_{\lambda \lambda '} f^\lambda(E_\nu),
$
where in this case $\nu$ ($\nu'$) represent energy eigenstates with energy $E_\nu$ ($E_{\nu'}$), $f^\lambda(E) = 1/(1+e^{\beta^\lambda (E - \mu^\lambda)})$ 
is the Fermi distribution function, $\beta^\lambda$ is the inverse temperature and  $\mu^\lambda$ is the chemical potential of lead $\lambda$. 

Now we substitute $\hat c_x(t)$ in Eq. (\ref{eq:noise_definition}) and evaluate the expectation values. It is straighforward but lengthy to show that in our system Wick's theorem remains valid for the time dependent creation and annihilation operators in the system, 
\begin{widetext}
\begin{equation} 
\langle
\hat c_{x}^{ \dagger}(t_1)
\hat c_{y}^{ \dagger}(t_2)
\hat c_{z}(t_3)
\hat c_{l}(t_4) \rangle    
  = \langle
\hat c_{x}^{ \dagger}(t_1)
\hat c_{l}(t_4) \rangle 
\langle
\hat c_{y}^{ \dagger}(t_2)
\hat c_{z}(t_3) \rangle  -
\langle
\hat c_{x}^{ \dagger}(t_1)
\hat c_{z}(t_3) \rangle
\langle
\hat c_{y}^{ \dagger}(t_2)
\hat c_{l}(t_4) \rangle.
\end{equation}
Using the above relation, Eq.~(\ref{eq:noise_definition}) simplifies to
$
S_{x y} (t,s) =  \re   \langle   \hat c_x (t) \hat c_{y}^\dagger (s)  \rangle \langle \hat c_x^{\dagger}(t) \hat c_y(s) \rangle.
$
The expectation values $\langle \hat c^\dagger \hat c \rangle$  and $\langle \hat c \hat c^\dagger\rangle$ are evaluated using Eqs.~(\ref{eq:solns}). We obtain 
$
\langle \hat c^\dagger_x(t) \hat c_y(s) \rangle 
= (1/\pi) \sum_{\lambda} \int d \omega e^{i  \omega( t - s) }
f^\lambda(\omega) \langle y |
G(s,\omega) \Gamma^{\lambda } (\omega)  
G^\dagger(t,\omega) | x \rangle 
$ and
$
\langle   \hat c_x (t) \hat c_{y}^\dagger(s)  \rangle 
= (1/\pi) \sum_{\lambda } \int d \omega  e^{-i\omega (t 	- s)} \bar f^\lambda(\omega)
\langle x | 
G(t,\omega) 
\Gamma^{\lambda}(\omega) G^\dagger(s,\omega) | y \rangle,\;
$
where $\bar f^\lambda = 1-f^\lambda$. 
 These results lead to the time-dependent voltage noise
\begin{align}
S_{x y} (t,s) & = \frac{1}{\pi^2}\re
\sum_{\lambda \kappa} \int d \omega d \omega' e^{i  (\omega-\omega')( t - s) }
f^\lambda(\omega) \langle y | 
G(s,\omega) \Gamma^{\lambda } (\omega) 
G^\dagger(t,\omega) | x \rangle 
  \bar f^\kappa(\omega') \langle x | 
G(t,\omega')  \Gamma^{\kappa}(\omega')
G^\dagger(s,\omega') | y \rangle.
\end{align}

Taking the Fourier transform, we obtain the Floquet noise power spectrum as shown
in the main text
\begin{align}
S^{(m)}_{xy} (\omega) 
&= 2\pi \re \sum_{{\substack{kln \\ \lambda \kappa}}}
\int d \nu
\bra{y}W^{\lambda}_{k+n,n+l}(\nu)\ket{x} f^\lambda(\nu) 
 {\bra{x}W^{\kappa}_{m+l,k}( \nu+\omega+n\Omega)\ket{y}}
\bar f^{\kappa}( \nu+\omega+n\Omega),	
\label{eq:S}
\end{align}
\end{widetext}
where $k,l,n,$ are integers,
the Fermi distribution
$f^\lambda(\omega) = \left(1+\exp[\beta^\lambda(\omega-\mu^\lambda)]\right)^{-1}$ of lead $\lambda$ with chemical potential $\mu^\lambda$ and inverse-temperature $\beta^\lambda$, $\bar f^\lambda = 1-f^\lambda$, and $W^{\lambda}_{n,m}(\omega) = G^{(n)}(\omega) \Gamma^\lambda(\omega) G^{(m) \dagger}(\omega) = W^{\lambda\dagger}_{m,n}(\omega)$, with  $\Gamma^\lambda$ the tunnel coupling matrix. For the one-dimensional model considered in the main text, 
	$\langle x | \Gamma^\lambda | y \rangle= \delta_{xy} \delta_{x \lambda} \gamma^\lambda $ in position representation, and the parameter $\gamma^\lambda$ characterizes the lead-system coupling strength.

\section{Weak coupling limit}
In this section we present the voltage noise power spectrum in the wide-band weak-coupling limit.
In the wide-band limit, the calculation of the Green's function is equivalent to solving the Floquet equation \cite{kohler2005}
\begin{equation}
\left[ w(t)-i\Gamma-i d/dt \right] | u_\alpha(t) \rangle = z_\alpha
| u_\alpha (t) \rangle
\end{equation}
where $\Gamma= \sum_\lambda \Gamma^\lambda$ is constant and in the weak coupling limit is treated
as a perturbation. To lowest order in $\Gamma=\sum_\lambda\Gamma^\lambda$, 
$z_\alpha = \epsilon_\alpha + i \gamma_\alpha$,
$ \gamma_\alpha = -\sum_k\bra{u_\alpha^{(k)}}\Gamma\ket{u_\alpha^{(k)}}$
and the Floquet Green's function has spectral representation
\begin{equation}\label{eq:Gnw}
G^{(n)}(\omega) =\sum_{k \alpha} \frac{\ket{u^{(k+n)}_\alpha}\bra{\bar u^{(k)}_\alpha}}{\omega - (\epsilon_\alpha + k \Omega + i\gamma_\alpha	 )}.
\end{equation}
In this limit, the propagators $W^\lambda$ are given by
\begin{align} \nonumber
W^\lambda_{n',n}(\omega)
& \approx \frac{1}{\pi}
\sum_{\alpha k} \Upsilon^{\lambda(k)}_\alpha
| u^{(n'+k)}_{\alpha} \rangle  \langle u^{(n+k)}_{\alpha} | \delta_{\gamma_\alpha}(\omega - \epsilon_\alpha - k \Omega),
\end{align}
where
$
\Upsilon^{\lambda(k)}_\alpha = \pi \bra{u_\alpha^{(k)}} \Gamma^\lambda\ket{u_\alpha^{(k)}}/\gamma_\alpha
$
and $\delta_\varepsilon(x) = (\varepsilon/\pi)/(x^2 + \varepsilon^2)$.
Substituting the above approximation for $W^\lambda_{n',n}(\omega)$  in Eq. (\ref{eq:S}) we obtain
the power spectrum in the weak-coupling limit
\begin{widetext}
\begin{align}
S^{(m)}_{x y} (\omega) &\approx
\pi \re \sum_{{\substack{k l n \\ \lambda \kappa \alpha\beta}}}
\Upsilon^{{\lambda} (k)}_\alpha \Upsilon^{{\kappa} (l)}_\beta
M_{\alpha\beta,xy}^{(m,n)} \left[ f^\lambda( \epsilon_\alpha+ k\Omega)  \right.
\bar f^{\kappa}(\omega+\epsilon_\alpha+ (n+l) \Omega ) +
\bar f^{\kappa}(\epsilon_\beta + l \Omega ) \nonumber \\
& \left.  f^\lambda( -\omega + \epsilon_\beta - (n-k)\Omega) \right]
\delta_{\gamma_{\alpha\beta}}(\omega +\epsilon_\alpha- \epsilon_\beta   + n \Omega )
\end{align}
with the matrix elements
$
M_{\alpha\beta,xy}^{(m,n)} = \sum_{q,p}
\inner{u^{(n+q)}_\alpha}{x}
\inner{x}{u^{(m+q)}_\beta}
\inner{u^{(p)}_\beta}{y}
\inner{y}{u^{(n+p)}_\alpha},
$
and
$\gamma_{\alpha\beta} \approx \max(\gamma_\alpha,\gamma_\beta) + \gamma_0$
the small broadening entering the Lorentzian
$ \delta_{\varepsilon}(z) = 
(\varepsilon/\pi)/(z^2+\varepsilon^2)$. When the  spectral broadening induced by the
coupling with the leads is larger than the temperature, the noise power spectrum
takes the more physically transparent form
\begin{align}
S^{(m)}_{x y} (\omega) &\approx
2\pi \re \sum_{{\substack{k l n \\ \lambda \kappa \alpha\beta}}}
\Upsilon^{{\lambda} (k)}_\alpha \Upsilon^{{\kappa} (l)}_\beta
f^\lambda( \epsilon_\alpha + k\Omega) \bar f^{\kappa}(\epsilon_\beta + l \Omega )
M_{\alpha\beta,xy}^{(m,n)}
\delta_{\gamma_{\alpha\beta}}(\omega +\epsilon_\alpha- \epsilon_\beta   + n \Omega )
\label{eq:weak}
\end{align}
presented in the main text. 

\begin{figure*}[t]
	\centering
	\includegraphics[width=16.5cm]{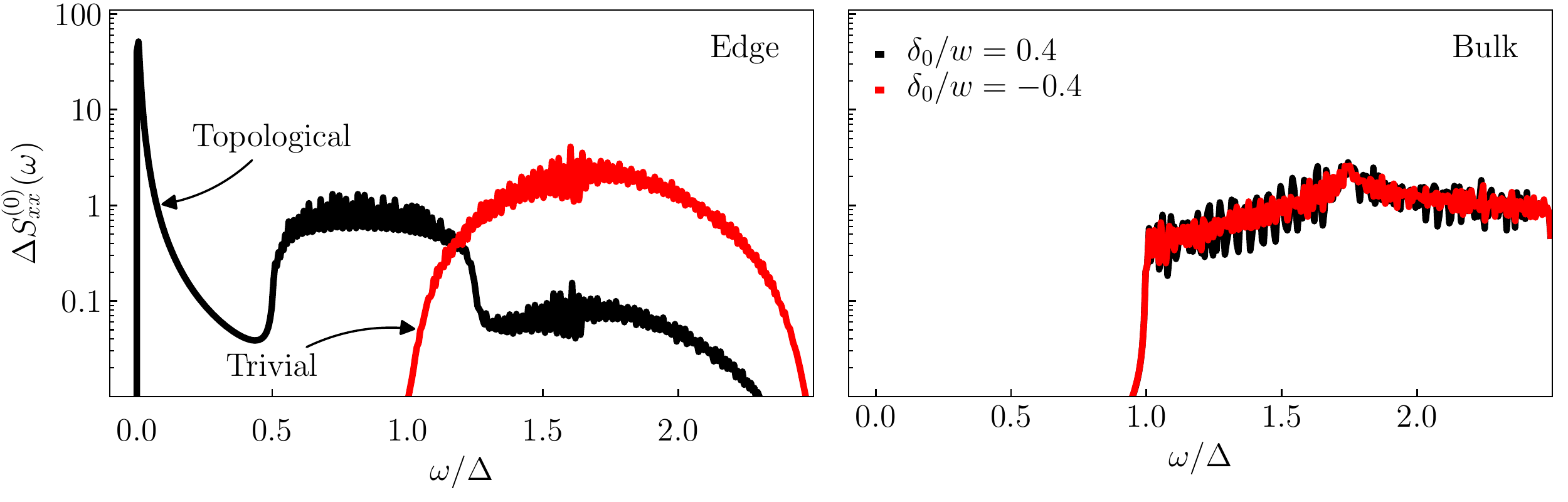}
	\caption{ Voltage noise spectrum at the (left) edge $x=1$ and bulk
		(right) $x=50$ as a function of the frequency $\omega/\Delta$
		for a static SSH model with $N=100$ sites. The black and red curves correspond to the topological
		($\delta_0/w = 0.4$), and trivial ($\delta_0/w = -0.4$) regimes respectively, where $w$ is the hopping amplitude. The gap in the energy spectrum is
		$\Delta/w \approx 1.6 $, $w\beta^\lambda = 10^{4}$, $\mu^\lambda/w = 0$, and $\gamma^\lambda/w = 10^{-2}$.
	}
	\label{fig:wc_1}
\end{figure*}

\vspace{-5mm}
\section{Static SSH model}

In this section, we present our results for a static SSH model \cite{su1979} . In the weak \-coupling limit, the voltage noise is given by
\begin{equation}
S^{(0)}_{xy}(\omega) \approx 2\pi \re \sum_{\lambda\kappa\alpha\beta} \Upsilon_\alpha^\lambda \Upsilon_\beta^\kappa P_{\alpha,yx} P_{\beta,xy} f^\lambda(E_\alpha) \bar f^\kappa(E_\beta) \delta(\omega+E_\alpha-E_\beta),
\end{equation}
\end{widetext}
where $P_{\alpha,xy} = 
\inner{x}{u_\alpha}\inner{u_\alpha}{y}$ are the matrix elements of the projector to eigenstate $| u_\alpha \rangle $ with energy $E_\alpha$, 
$
\Upsilon^{\lambda}_\alpha = \pi \bra{u_\alpha} \Gamma^\lambda\ket{u_\alpha}/\bra{u_\alpha}\Gamma\ket{u_\alpha}
$,
and
$\langle x | \Gamma^\lambda | y \rangle= \delta_{xy} \delta_{x \lambda} \gamma^\lambda$. 
In figure \ref{fig:wc_1} we plot the noise power spectrum
at the  edge of a sample (left panel) with $N=100$ sites for trivial, and topological regimes.
The frequency is normalized to the energy spectrum gap $\Delta$. In the topological regime, $ S^{(0)}_{xx} (\omega) $ at the edge has a peak at 
$\omega/\Delta \approx 0$, stemming from particle-hole pair exchanges 
between the localized edge states and the leads. As expected, this sharp feature is not present in the trivial regime. In contrast, the noise in the bulk (right panel) is similar in both regimes. The spectrum in the bulk shows a slight enhancement around $\omega/\Delta \approx 1.75 $. This feature arises from transitions between the bottom of the spectrum and states close to the gap edge, where  the density of states is relatively high. 

\end{document}